\documentclass[pdflatex,sn-nature]{sn-jnl}

\usepackage{graphicx}
\usepackage{amsmath,amssymb,amsfonts}
\usepackage{setspace}
\usepackage{lineno}
\usepackage{siunitx}

\graphicspath{{figures/}{./}}

\unnumbered

\begin{document}

\title[Article Title]{Extinction spectroscopy of a single colloidal perovskite quantum dot coupled to a tunable microcavity}

\author[1]{\fnm{Marina} \sur{Cagnon-Trouche}}

\author[1]{\fnm{Zakaria} \sur{Said}}

\author[2]{\fnm{Jakob} \sur{Reichel}}

\author[1]{\fnm{Christophe} \sur{Voisin}}

\author*[1]{\fnm{Yannick} \sur{Chassagneux}}
\email{yannick.chassagneux@phys.ens.fr}

\author*[1,3]{\fnm{Carole} \sur{Diederichs}}
\email{carole.diederichs@phys.ens.fr}

\affil[1]{\orgdiv{Laboratoire de Physique de l’Ecole normale supérieure}, \orgname{ENS, Université PSL, CNRS, Sorbonne Université, Université Paris Cité}, \orgaddress{\street{24 rue Lhomond}, \city{Paris}, \postcode{F-75005}, \country{France}}}

\affil[2]{\orgdiv{Laboratoire Kastler Brossel}, \orgname{Sorbonne Université, CNRS, ENS - Université PSL, Collège de France}, \orgaddress{\city{Paris} \postcode{F-75252},  \country{France}}}

\affil[3]{\orgname{Institut Universitaire de France}, \orgaddress{\city{Paris}, \postcode{75231}, \country{France}}}

\abstract{\noindent The reduction of the transmission of a macroscopic light beam by a single quantum emitter is a direct signature of efficient light-matter interaction. Pushed toward unity, this extinction enables giant single-photon nonlinearities for photonic quantum information processing. While near-unity values have been reached with single molecules, epitaxial quantum dots and colour centres in diamond coupled to optical microcavities, such an effect has remained largely unexplored for colloidal quantum dots. Here, we perform resonant transmission spectroscopy on a single CsPbBr$_3$ perovskite quantum dot deterministically coupled to a tunable open fibered Fabry-Pérot microcavity at \SI{10}{K}. Using a low-power broadband beam, we resolve an extinction dip for each line of the exciton fine structure, which grows up to about 10\% when reducing the cavity mode volume. The extracted vacuum Rabi couplings are consistent with those deduced on the same quantum dot from its cavity-modified emission spectrum and Purcell-accelerated decay.} 


\maketitle
%\doublespacing
%\linenumbers

% ===========================================================================
\section{Introduction}
% ===========================================================================

Controlling the interaction between a single quantum emitter and light is a key challenge of quantum optics. The minimal building block of such an interaction is the coupling of an individual two-level system to a single mode of the electromagnetic field. The reduction of light transmission, known as extinction, is the most direct probe of this fundamental interaction. It relies on the interference, in the forward direction, between the transmitted beam and the field coherently scattered by the emitter. For an ideal two-level system well below saturation and limited solely by its radiative lifetime (i.e. characterised by $T_2=2T_1$, with $T_1$ the lifetime and $T_2$ the coherence time), the extinction cross section reaches $3\lambda^2/2\pi$ at resonance and is thus set by the transition wavelength alone. With an optimally shaped wavefront, unity extinction can in principle be reached, enabling single-photon nonlinearities without any cavity \cite{Zumofen2008} and paving the way for quantum information processing. In practice, however, resolving an extinction dip has long been an experimental challenge, and typical values of the order of 10\% could only be measured by resonant laser transmission spectroscopy with a tightly focused beam, in single trapped atoms \cite{Tey2008}, epitaxial quantum dots \cite{Vamivakas2007} and molecules \cite{Pototschnig2011}. 

In the case of solid-state two-level systems, apart from the challenge of coupling light to an emitter that is usually embedded in a high-refractive-index matrix \cite{Tran2017}, the main limitation is set by the emitter itself. Pure dephasing reduces the fraction of coherently scattered light, given by $T_2/(2T_1)$, and non-radiative decay channels lower the radiative quantum yield, both effects bringing the extinction cross section far below its ideal value. Coupling the emitter to an optical cavity relaxes these limitations: in a simplified picture, the interaction is enhanced by the number of round trips of light in the cavity. In the framework of cavity quantum electrodynamics (cQED), the extinction is governed by the cooperativity $\mathcal{C}$, a dimensionless figure of merit that compares the coherent emitter-cavity coupling with the cavity losses and the emitter decoherence. A small mode volume, which enhances light-matter coupling, and a high quality factor, which reduces the cavity losses, can thus raise the cooperativity enough to counterbalance pure dephasing, so that the extinction approaches unity for cooperativities well above one. Near-unity extinction has been achieved in this way for single molecules \cite{Wang2019}, epitaxial quantum dots \cite{Tomm2024} and tin-vacancy centres in diamond \cite{Codreanu2026,Koster2026} coupled to optical microcavities. Despite this progress, a sizeable resonant extinction had not, to our knowledge, been observed in transmission for an individual colloidal quantum dot, although it would bring single-photon nonlinearities to a scalable, solution-processed material platform. 

Colloidal quantum dots have indeed long been limited by modest radiative quantum yields and large pure dephasing. Lead halide perovskite quantum dots (pQDs) have recently changed this picture. Their radiative quantum yield can reach up to $\sim$80\% at the single particle level, as shown recently on a large number of individual pQDs \cite{Gruber2026}, with an emission showing very limited blinking \cite{Raino2016}. The measurement of 50\% photon indistinguishability \cite{Kaplan2023}, which was unprecedented for colloidal emitters, also indicates that pure dephasing can be moderate, opening the way to advanced quantum optics studies with pQDs. Since then, single pQDs have been coupled to circular Bragg gratings \cite{Jun2024}, plasmonic cavities \cite{Li2026,Liao2026} and open Fabry-Pérot microcavities \cite{Oddi2026,Said2026}, where a Purcell acceleration of their spontaneous emission has been observed. The latter open microcavity geometry is particularly suited to quantitative cQED studies since the emitter can be deterministically coupled, both spatially and spectrally, to the cavity mode, and the mode volume can be finely tuned. This allowed us to measure the vacuum Rabi coupling and the instantaneous linewidth (i.e. the linewidth free of spectral diffusion) of a single CsPbBr$_3$ pQD from the cavity-induced modification of its emission spectrum and the Purcell acceleration of its decay measured by time-resolved photoluminescence \cite{Said2026}.

Beyond emission, this well characterised pQD-cavity system opens the way to other fundamental cQED studies. Here, we use it to probe directly the extinction of a single pQD by resonant transmission spectroscopy. Such measurements usually require a narrow-linewidth laser scanned across the resonance of an actively stabilised cavity. In our open cavity, the residual mechanical vibrations can be exploited to widen the spectral window: during each acquisition, they randomly sweep the cavity resonance across the excitonic line, so that a single transmission spectrum of a low-power broadband beam directly resolves the extinction dips. By tuning the longitudinal mode order, we show that the extinction increases as the mode volume decreases, as a consequence of an enhanced vacuum Rabi coupling, reaching unpolarised extinctions of 10\% and 7\% for the two fine structure lines at the smallest mode volume. Furthermore, the analysis of these transmission dips in the input-output framework yields vacuum Rabi coupling values consistent with those deduced from our previous independent emission study of the same pQD.

% ===========================================================================
\section{Results}
% ===========================================================================

\subsection*{Principle of the cavity extinction measurement}

\begin{figure}[htbp]
    \centering
    \includegraphics[width=\linewidth]{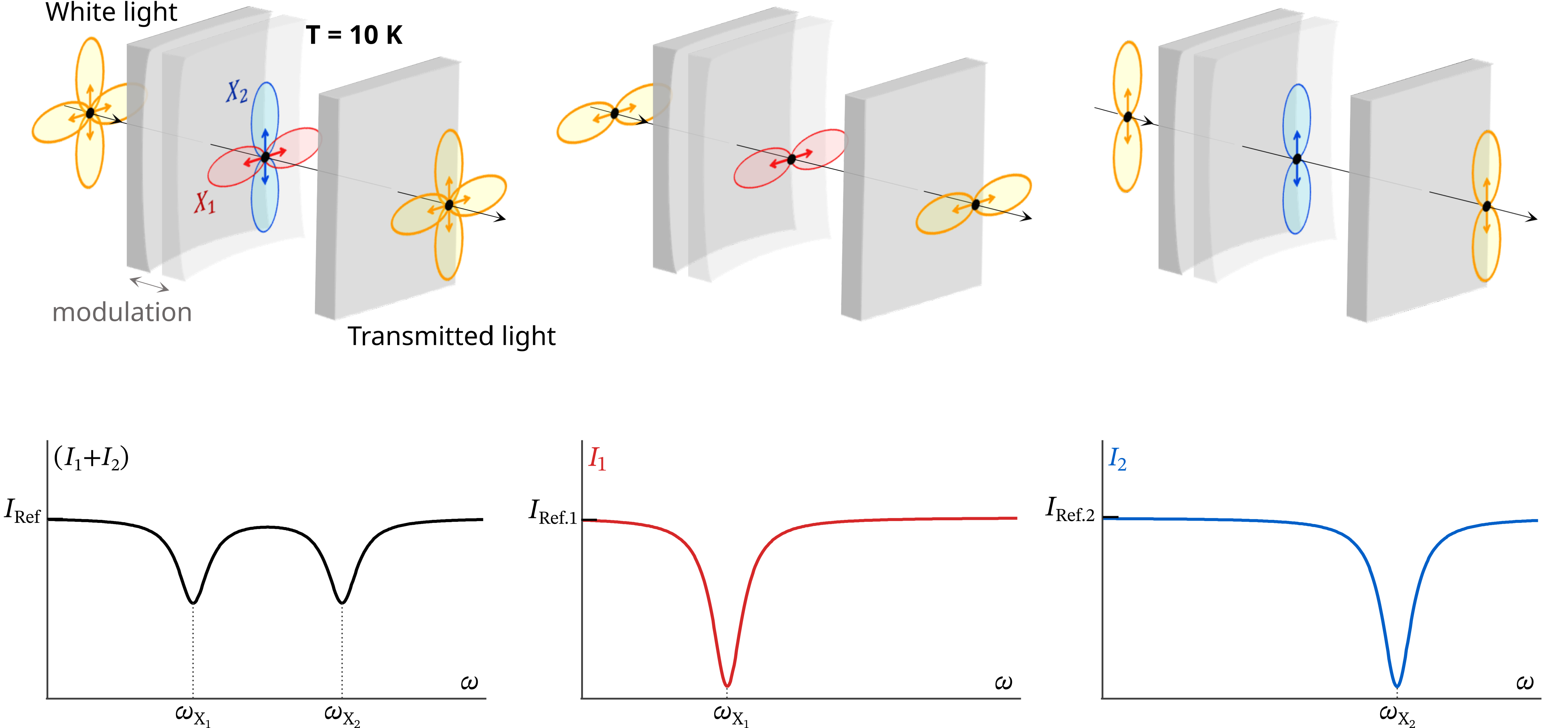}
    \caption{\textbf{Principle of the extinction measurement of a single perovskite quantum dot coupled to an open microcavity.} Top: a broadband white light beam probes a single pQD coupled to an open fibered Fabry-Pérot microcavity at $10$~\si{\kelvin}. The cavity length is modulated by residual mechanical vibrations (sub-kHz), sweeping the cavity resonance across each line of the excitonic doublet, so that a single \SI{5}{s} acquisition samples the whole transition. The pQD exciton fine structure is represented by two orthogonally polarised dipoles, $X_1$ (red) and $X_2$ (blue), each coupled to the co-polarised cavity mode. Bottom: schematic transmitted spectra. On resonance, each excitonic line reduces the intensity transmitted through the planar mirror in its own polarisation (red and blue spectra). Without polarisation analysis, the transmitted intensity (black spectrum) shows two dips, each about half as deep as in a polarisation-resolved measurement.}
    \label{fig:principle}
\end{figure}

We study a single CsPbBr$_3$ perovskite quantum dot (pQD) presenting an excitonic doublet and deterministically coupled to a tunable fibered Fabry-Pérot microcavity operated at \SI{10}{K}. This is the same pQD and cavity platform that we recently used to extract key cQED figures of merit \cite{Said2026}, enabling a direct and quantitative comparison between emission and extinction approaches. Fig.~\ref{fig:principle} summarises the principle of the extinction measurement. A low-power broadband unpolarised white light beam is injected through the fibre mirror into the microcavity. The length of the cavity, and hence the frequency of the longitudinal mode, is modulated by residual mechanical vibrations (sub-kHz), sweeping the mode back and forth across the excitonic transitions of the pQD. The intrinsic cavity linewidth, $\hbar\kappa\approx\SI{100}{\micro eV}$, is narrower than each exciton line of instantaneous linewidth $\hbar(\gamma+\gamma^*)\approx\SI{250}{\micro eV}$, i.e. the linewidth free of spectral diffusion \cite{Said2026}. Hence, a static cavity would probe a fraction of the lineshape only. The vibrations, leading to a Gaussian broadening of standard deviation $\sigma\approx 2$-$\SI{3}{meV}$ (FWHM of $\approx 5$-$\SI{7}{meV}$), instead average over the full excitonic transition.

The transmitted light is collected through the planar mirror, carrying the fingerprint of the pQD as dips in the transmission spectrum. At cryogenic temperature, the exciton fine structure consists of a doublet of cross-linearly polarised lines, $X_1$ and $X_2$, split by $\SI{700}{\micro eV}$ (Fig.~\ref{fig:transmission}b and Supplementary Fig.~S1), which we model as two orthogonal dipoles \cite{Amara2023}. The Fabry-Pérot cavity supports two orthogonally polarised modes whose splitting is small compared to the cavity linewidth. They can therefore be considered degenerate and described in the basis defined by the pQD dipole axes, so that each excitonic line couples to a single co-polarised mode, as depicted in Fig.~\ref{fig:principle}. As neither excitation nor detection is polarisation resolved in our experiment, the reference transmission, i.e. the cavity transmission without the contribution of the pQD, contains both cavity modes, whereas each excitonic line attenuates only its co-polarised mode. Each line ($X_1$ and $X_2$) thus produces a dip at its own frequency ($\omega_{X_1}$ and $\omega_{X_2}$), weighted by the fraction of the reference carried by the corresponding polarisation, i.e. about one half for unpolarised light. In the full transmission spectrum (in black in Fig.~\ref{fig:principle}), the depth of each dip, i.e. the unpolarised extinction of the corresponding line, is therefore about half of that obtained in a polarisation-resolved measurement (Supplementary Note 1).

\subsection*{Transmission spectrum}

\begin{figure}[htbp]
    \centering
    \includegraphics{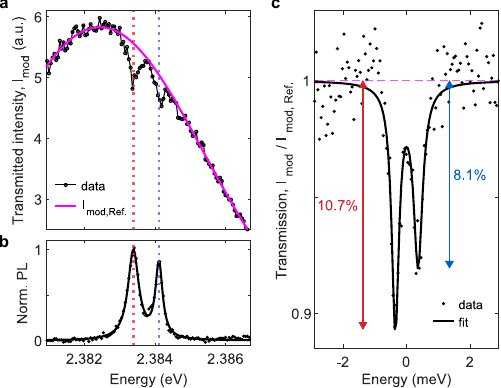}
    \caption{\textbf{Transmission of a single CsPbBr$_3$ quantum dot in cavity.} (a) Transmitted intensity $I_\text{mod}$ of the cavity tuned to a single pQD (mode $p=15$). The baseline $I_\mathrm{mod,Ref}$ (magenta) is obtained from a Gaussian fit excluding a \SI{2}{meV}-wide window centred on the excitonic doublet. (b) Photoluminescence spectrum of the pQD coupled to the cavity; the dotted lines indicate the positions of $X_1$ (red) and $X_2$ (blue). (c) Normalised transmission $I_\text{mod}/I_\text{mod,Ref}$ (dots), fitted by a sum of two Lorentzians (solid line). The arrows indicate the minima, 10.7\% for $X_1$ and 8.1\% for $X_2$ below the reference (magenta dashed line), slightly deeper than the unpolarised extinctions given by the Lorentzian amplitudes, $a_1=10.5\%$ and $a_2=7.7\%$.}
    \label{fig:transmission}
\end{figure}

With the cavity resonance tuned on average to the pQD and swept by the vibrations, we record the transmitted intensity $I_\text{mod}$ on a spectrometer. A representative spectrum is shown in Fig.~\ref{fig:transmission}a, where a broad Gaussian-like cavity transmission envelope shows two shallow dips that coincide with the $X_1$ and $X_2$ lines measured by photoluminescence (Fig.~\ref{fig:transmission}b). To convert the raw intensity into a transmission spectrum, we normalise $I_\text{mod}$ by a reference $I_\mathrm{mod,Ref}$ corresponding to the intensity transmitted by the cavity in the absence of the pQD. Rather than measuring this reference in a separate acquisition, which would not be reliable as the transmitted intensity depends sensitively on the local measurement conditions and can vary between acquisitions, we extract $I_\mathrm{mod,Ref}$ directly from the same transmission spectrum, by a Gaussian fit of the envelope excluding a \SI{2}{meV}-wide window centred on the excitonic doublet (magenta line in Fig.~\ref{fig:transmission}a; see Methods). For Fig.~\ref{fig:transmission}a, we chose an acquisition in which the exciton lines lie on the flank of the vibration distribution: the maximum of the Gaussian envelope then falls outside the excluded window, which better constrains the fit and hence the reference. Acquisitions with the lines close to the centre of the distribution show the same two dips (Supplementary Fig.~S3), and, as shown below, the unpolarised extinction does not depend on where the emitter lies within the distribution. The normalised transmission $I_\text{mod}/I_\mathrm{mod,Ref}$ then reveals two distinct minima, which we fit with a sum of two Lorentzians centred on $X_1$ and $X_2$ (Fig.~\ref{fig:transmission}c; Eq.~\eqref{eq:dip_fitting} in Methods). We characterise the unpolarised extinction of each line by the amplitude $a_i$ of its Lorentzian. As the two lines are split by only about 2.5 times their linewidth, the normalised transmission at the centre of each line also contains the tail of the other line, and is therefore slightly deeper than $a_i$ (Supplementary Fig.~S2). For the mode shown in Fig.~\ref{fig:transmission} ($p=15$), the minima reach 10.7\% for $X_1$ and 8.1\% for $X_2$ (arrows in Fig.~\ref{fig:transmission}c), corresponding to unpolarised extinctions $a_1=10.5\%$ and $a_2=7.7\%$. Repeated acquisitions at a given mode yield consistent values: ten acquisitions recorded on different days at $p=15$ give mean unpolarised extinctions $\langle a_1\rangle=9.5\%$ and $\langle a_2\rangle=7.1\%$ (Supplementary Fig.~S4 and Supplementary Note 2), i.e. about 19\% and 14\% in a polarisation-resolved measurement.

\subsection*{Extinction versus mode volume and vacuum Rabi coupling} 

\begin{figure}[htbp]
    \centering
    \includegraphics{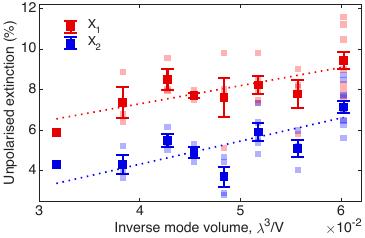}
    \caption{\textbf{Unpolarised extinction of the two fine structure lines of a single CsPbBr$_3$ quantum dot in cavity.} Unpolarised extinction of lines $X_1$ (red) and $X_2$ (blue), given by the Lorentzian amplitudes $a_1$ and $a_2$, as a function of the inverse cavity mode volume $\lambda^3/V$, for the longitudinal modes $p = 15$ to 26. Light symbols are derived from individual transmission measurements; dark symbols are the average for each mode, the error bars showing the corresponding spread. The dotted lines are guides for the eye. }
    \label{fig:extinction}
\end{figure}

To study the effect of the cavity mode volume, we repeated the measurement across the longitudinal modes $p = 15$ to 26, and extracted for each acquisition the unpolarised extinctions $a_1$ and $a_2$ of $X_1$ and $X_2$. As shown in Fig.~\ref{fig:extinction}, despite the dispersion between acquisitions, the unpolarised extinction of both lines increases overall as the mode volume decreases. This increase with $\lambda^3/V$ is the signature of light-matter coupling in the cQED regime.

\begin{figure}[htbp]
    \centering
    \includegraphics{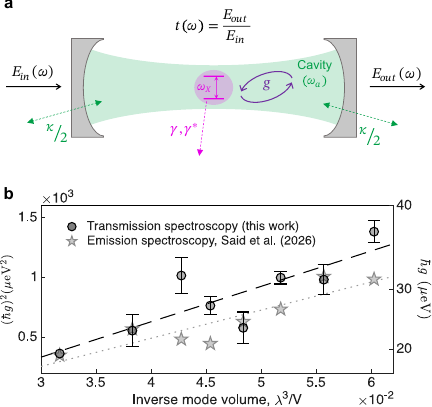}
    \caption{\textbf{Vacuum Rabi coupling of a single CsPbBr$_3$ quantum dot in cavity.} (a) Sketch of the input-output model: a dissipative two-level system (magenta) at frequency $\omega_X$ with instantaneous linewidth $\gamma+\gamma^*$ couples with a vacuum Rabi coupling $g$ to a cavity mode of frequency $\omega_a$ and losses $\kappa$ ($\kappa/2$ through each mirror). The incident field $E_\text{in}$ is injected through one mirror and the transmitted field $E_{\mathrm{out}}$ is collected through the other. Their ratio defines the transmission amplitude $t$ [Eq.~\eqref{eq:transmittivity}]. (b) Squared vacuum Rabi coupling $(\hbar g)^2$, averaged over the two fine structure lines and extracted from the total area of the two extinction dips (black dots, this work), as a function of the inverse mode volume $\lambda^3/V$. Grey stars show the values obtained independently on the same pQD from its cavity-modified emission spectrum and Purcell acceleration \cite{Said2026}. Dashed and dotted lines are linear fits.}
    \label{fig:rabi}
\end{figure}

To relate this increase to the vacuum Rabi coupling $g$, we model the modulated transmission. The transmission amplitude of a dissipative two-level system coupled to a cavity mode is derived from the Gardiner \& Collett input-output formalism \cite{Gardiner1985}, where the radiative decay $\gamma$ and pure dephasing $\gamma^*$ are introduced phenomenologically \cite{Auffeves2010}. Fig.~\ref{fig:rabi}a illustrates the main parameters of the model leading to the transmission amplitude:
\begin{equation}
  t(\omega) = \frac{E_\text{out}}{E_\text{in}} =
  \frac{\dfrac{\kappa}{2}}{\dfrac{\kappa}{2} - i(\omega - \omega_a)
    + \dfrac{g^2}{\dfrac{\gamma+\gamma^*}{2} - i(\omega - \omega_X)}} ,
  \label{eq:transmittivity}
\end{equation}
where $E_{\mathrm{in}}$ and $E_{\mathrm{out}}$ are the incident and transmitted fields injected and collected through opposite mirrors, respectively, $\kappa$ is the total cavity loss rate ($\kappa/2$ through each mirror), $\omega_a$ and $\omega_X$ are the cavity mode and emitter angular frequencies, $g$ is the vacuum Rabi coupling, and $\hbar(\gamma+\gamma^*)$ corresponds to the emitter instantaneous linewidth. 

The mechanical vibrations of the open microcavity make the mode frequency $\omega_a$ fluctuate around a mean value $\bar\omega_a$ with a Gaussian distribution $P(\omega_a)$ of standard deviation $\sigma$. As the two cavity modes are orthogonally polarised, they do not interfere, and the measured intensity is the sum of the intensities they transmit, each averaged over the vibrations:
\begin{equation}
I_\text{mod,tot}(\omega) \propto \int \left(|t_{X_1}(\omega,\omega_a)|^2 + |t_{X_2}(\omega,\omega_a)|^2 \right) P(\omega_a)\,\mathrm{d}\omega_a ,
  \label{eq:Ttot_incoherent}
\end{equation}
where $t_{X_i}$ is the transmission amplitude of the mode coupled to line $X_i$.
In our experimental regime where the vibration amplitude dominates all spectral widths, each extinction dip reduces to a Lorentzian that does not depend on the offset $\bar\omega_a-\omega_X$ between the mean cavity frequency and the emitter frequency (Methods and Supplementary Fig.~S6). This is essential in practice since this offset varies from one acquisition to the next. For a co-polarised probe, the dip is centred at the emitter frequency $\omega_X$, with an amplitude $\mathcal{C}/(1+\mathcal{C})$ and a FWHM $(\gamma+\gamma^*)\sqrt{1+\mathcal{C}}$, where
\begin{equation}
\mathcal{C}=\frac{4g^{2}}{\kappa(\gamma+\gamma^*)}
\label{eq:cooperativity}
\end{equation}
is the cooperativity of the line, which quantifies the strength of the emitter-cavity coupling relative to the combined cavity and emitter loss rates \cite{Auffeves2007}. Without polarisation analysis, each dip keeps the same FWHM and its amplitude, the unpolarised extinction, is reduced by half.

We estimate $g$ from the total area $\Lambda$ of the two dips in the normalised transmission, deduced from the two-Lorentzian fit (Supplementary Note 4). In the weak cooperativity limit ($\mathcal{C}\ll 1$), the area of a co-polarised dip reduces to $2\pi g^{2}/\kappa$, lifting the need for the difficult assessment of the emitter intrinsic (spectral diffusion free) linewidth required for other methods. As each line contributes half of this area, the total area is
\begin{equation}
    \Lambda \approx \frac{2\pi g^{2}}{\kappa},
\end{equation}
where $g^2$ is now the mean of the squared couplings of the two lines (Supplementary Note 4) and the cavity loss rate $\kappa$ is measured for each longitudinal mode. Figure~\ref{fig:rabi}b shows the resulting $(\hbar g)^{2}$ as a function of the inverse mode volume $\lambda^3/V$ (black dots). It increases linearly with $\lambda^3/V$, reaching $\hbar g\approx\SI{37}{\micro eV}$ at the smallest mode volume, which corresponds to a mean cooperativity $\mathcal{C}\approx0.2$. The $g$ values follow the same scaling as those obtained independently on the same pQD from its cavity-modified emission spectrum and Purcell acceleration (grey stars) \cite{Said2026}. The convergence of these two independent techniques, extinction and emission, supports the reliability of the extracted light-matter coupling.

\subsection*{Extinction linewidths}

The Lorentzian fits also give the dip linewidths $(\gamma+\gamma^*)\sqrt{1+\mathcal{C}}$. Once corrected for this slight cQED-induced broadening, they yield the emitter linewidth, which may include spectral diffusion broadening (Supplementary Note 2 and Supplementary Fig.~S5). Averaged over all modes, we obtain $262\pm9$~\si{\micro\electronvolt} for $X_1$ and $242\pm14$~\si{\micro\electronvolt} for $X_2$. For comparison, the free-space PL linewidths recorded on the same pQD by opening the cavity between extinction acquisitions are $311\pm4$~\si{\micro\electronvolt} for $X_1$ and $262\pm3$~\si{\micro\electronvolt} for $X_2$. Both lines are thus narrower in extinction than in PL, as expected from the contrasting excitation regimes: extinction probes the emitter on resonance at very low power, whereas PL requires a much higher off-resonant power, which may photogenerate charges around the pQD and broaden the line through spectral diffusion. More quantitatively, the $X_1$ line is 16\% narrower in extinction than in PL, a statistically significant difference according to the Student criterion ($t_s\simeq 4.98$; see Methods). The $X_2$ line also shows an 8\% reduction, but this difference is not statistically significant ($t_s\simeq 1.40$) and does not allow a firm conclusion. Overall, the linewidths extracted from extinction are consistent within their uncertainties with the instantaneous value of $\hbar(\gamma+\gamma^*)=\SI{250}{\micro eV}$ obtained independently for this pQD \cite{Said2026}, showing that spectral diffusion contributes little to the extinction linewidths.

% ===========================================================================
\section{Discussion}
% ===========================================================================

We have resolved the resonant extinction of a light beam by a single colloidal quantum dot. For a CsPbBr$_3$ pQD coupled to a tunable open microcavity, the two fine structure lines reach unpolarised extinctions of about 10\% and 7\% at the smallest mode volume, corresponding to a mean cooperativity $\mathcal{C}\approx0.2$. Modelling the vibration-averaged transmission allows us to extract the vacuum Rabi coupling from the dip areas, and the values are consistent with those measured by emission spectroscopy on the same pQD \cite{Said2026}. Resonant transmission thus provides an independent and quantitative access to the light-matter coupling of a single colloidal emitter, and validates extinction as a quantitative probe of cQED in these systems. The measurement itself is also straightforward and robust: the residual vibrations of the open cavity sweep the cavity mode across the excitonic line within each acquisition and make the extracted extinction independent of the cavity-emitter average offset, so that cavity stabilisation is not required.

The magnitude of the extinction results from the competition between coherent coupling and decoherence. The pQD lifetime-limited linewidth $\hbar\gamma\approx\SI{10}{\micro eV}$, deduced from the free-space lifetime, is much smaller than the instanteneous linewidth $\hbar(\gamma+\gamma^*)\approx\SI{250}{\micro eV}$, so that only about 4\% of the scattered light is coherent. In free space, together with a finite radiative quantum yield, this would limit the extinction of a realistic focused beam to the $10^{-3}$ range, which is very challenging to resolve. Here, the cavity makes the extinction of the pQD measurable: the relevant figure of merit becomes the cooperativity $\mathcal{C}$, in which a large light-matter coupling can compensate for large dephasing. Our system nevertheless remains far from the regime $\mathcal{C}\gtrsim1$ required for near-unity extinction. Reaching it requires increasing $\mathcal{C}$ by improving either the cavity or the emitter. 

On the cavity side, smaller mode volumes \cite{Borel2023} and higher finesses \cite{Najer2019} would directly increase $\mathcal{C}$. On the emitter side, the cooperativity can also be raised through the properties of the nanocrystals. The vacuum Rabi coupling scales as $g\propto\sqrt{f/V}$, with $f$ the oscillator strength of the transition. In weakly confined CsPbBr$_3$ pQDs, $f$ is not fixed by the material but grows with the nanocrystal volume beyond the exciton Bohr radius, as evidenced by single-photon superradiance \cite{Zhu2024} and superabsorption \cite{Boehme2025}. This route is effective precisely because the emitter is limited by pure dephasing: when $\gamma^*\gg\gamma$, $\mathcal{C}\approx4g^2/(\kappa\gamma^*)$ grows linearly with $f$, whereas for a radiatively limited emitter the concomitant increase of $\gamma\propto f$ would cancel the gain. At fixed cavity and dephasing, a larger oscillator strength would therefore increase the cooperativity proportionally, while also improving the coherent fraction $\gamma/(\gamma+\gamma^*)$ of extinction. This gain is bounded, as the enhancement of $f$ is ultimately limited by the exciton coherence length \cite{Zhu2024}, but larger dots are readily synthesised with edge lengths up to \SI{30}{nm} \cite{Boehme2025}. Larger nanocrystals may also be less affected by pure dephasing. Part of the broadening may indeed arise from fluctuating charges in the pQD environment, to which larger dots, with a smaller surface-to-volume ratio, should be less sensitive. Operating closer to 4~\si{\kelvin} would further reduce the contribution of elastic phonon scattering, identified as the dominant dephasing mechanism in weakly confined CsPbBr$_3$ nanocrystals at cryogenic temperatures \cite{Sun2023}.

Beyond these improvements, resolving the extinction of a single colloidal quantum dot gives access to its coherent optical response, complementary to emission. The phase shift imprinted by the emitter on the transmitted beam could be measured by interference experiments, and the dependence of the extinction on the excitation power, or the photon statistics of the transmitted light, would probe the onset of single-photon nonlinearities. Such experiments would bring colloidal perovskite quantum dots a step closer to the coherent light-matter coupling targeted by quantum photonics.

% ===========================================================================
\section{Methods}
% ===========================================================================

\subsection*{Sample preparation}
CsPbBr$_3$ perovskite quantum dots were synthesised by the hot-injection method \cite{Protesescu2015} and, after dilution, embedded in a $\sim200$~\si{\nano\meter}-thick polystyrene matrix spin-coated on a high-reflectivity planar dielectric mirror, at a typical density of 0.01~\si{\micro\meter^{-2}}. The polystyrene matrix stabilises the emission and protects the pQDs from degradation. Full synthesis and sample preparation details are given in \cite{Said2026}.

\subsection*{Tunable fibered Fabry-Pérot microcavity}
The plano-concave open microcavity is composed of a planar substrate mirror and a concave dielectric mirror micro-machined at the tip of an optical fibre (radius of curvature $\approx10$~\si{\micro\meter}). Both mirrors are highly reflective at the pQD emission wavelength ($R\simeq 0.9995$ in the 475~\si{\nano\meter} to 585~\si{\nano\meter} range), yielding a finesse $\mathcal{F}\simeq1500$ and a quality factor $Q=25000-50000$ depending on the longitudinal mode order $p$. Spatial and spectral matching are achieved by nanopositioning the mirrors; the cavity can be opened and closed reversibly, allowing the same pQD to be studied in free space and in cavity while keeping the local environment unchanged \cite{Said2026}.

\subsection*{Extinction measurement}
All measurements were performed at 10~\si{\kelvin} in a closed-cycle helium cryostat. For the extinction measurement, a low-power broadband white-light source ($P\approx80$~\si{\nano\watt}) is injected through the fibre; the transmitted light is collected through the planar mirror and dispersed on a grating spectrometer, with a \SI{5}{s} integration time per spectrum. The cavity length, and hence the mode frequency, is modulated at sub-kHz by residual mechanical vibrations, which sweep the cavity across the excitonic lines within a single acquisition. Longitudinal modes $p=15$ to $26$ were investigated, spanning a range of mode volumes $V$.

\subsection*{Baseline extraction and dip fitting}
The transmission is obtained by normalising the modulated transmitted intensity $I_\text{mod}$, recorded with the cavity tuned to the pQD, by a reference $I_\mathrm{mod,Ref}$ extracted from the same spectrum. For the latter, a Gaussian is fitted to the cavity envelope after excluding a fixed \SI{2}{meV} window centred on the exciton doublet. This procedure accounts for the slow variations of the baseline between acquisitions. The normalised transmission is fitted by a sum of two Lorentzians,
\begin{equation}
\frac{I_\text{mod}(\omega)}{I_\mathrm{mod,Ref}(\omega)} = 1
-\frac{a_1}{1+\left(\dfrac{\omega-\omega_{X_1}}{\Gamma_1/2}\right)^{2}}
-\frac{a_2}{1+\left(\dfrac{\omega-\omega_{X_2}}{\Gamma_2/2}\right)^{2}} ,
\label{eq:dip_fitting}
\end{equation}
where $a_i$, $\omega_{X_i}$ and $\Gamma_i$ are the amplitude, centre frequency and FWHM of the dip of line $X_i$. The amplitude $a_i$ gives the unpolarised extinction of line $X_i$. The minimum of the normalised transmission at $\omega_{X_i}$ is slightly deeper than $a_i$, as it also contains the tail of the other line.

Statistical significance of the difference between the extinction and PL linewidths, $\Gamma_{\mathrm{ext}}$ and $\Gamma_{\mathrm{PL}}$ respectively, is assessed with a Student criterion $t_s=|\langle\Gamma_{\mathrm{ext}}\rangle-\langle\Gamma_{\mathrm{PL}}\rangle|/\sqrt{\sigma_{\mathrm{ext}}^{2}+\sigma_{\mathrm{PL}}^{2}}$, where $\sigma_{\mathrm{ext}}$ and $\sigma_{\mathrm{PL}}$ are the standard errors of the means and $t_s>2$ indicates statistically distinct means.

\subsection*{Model of the modulated transmission}
The transmission amplitude of Eq.~\eqref{eq:transmittivity} is derived from the Gardiner \& Collett input-output formalism \cite{Gardiner1985}, with the radiative decay $\gamma$ and pure dephasing $\gamma^{*}$ introduced phenomenologically \cite{Auffeves2010}. The cavity length modulation is described by a normalised Gaussian distribution of the mode frequency,
\begin{equation}
    P(\omega_a)=\frac{1}{\sigma\sqrt{2\pi}}\exp\!\left[-\frac{(\omega_a-\bar\omega_a)^2}{2\sigma^2}\right],
  \label{eq:distribution_vibrations}
\end{equation}
where $\sigma$ is the standard deviation of the frequency fluctuations. Their FWHM, $2\sqrt{2\ln2}\,\sigma\approx\SI{6}{meV}$, is the width of the Gaussian envelope in Fig.~\ref{fig:transmission}a. For each polarisation, the measured intensity is the average of $|t|^{2}$ weighted by $P(\omega_a)$, and the two polarisations add up in intensity [Eq.~\eqref{eq:Ttot_incoherent}]. For an empty cavity ($g=0$), $I_\text{mod}$ is the convolution of a Lorentzian empty cavity (FWHM $\kappa$) and the Gaussian (standard deviation $\sigma$), leading to a Voigt profile. In the regime $\sigma \gg \kappa$ relevant here ($\sigma\approx\SI{2.5}{meV}$, $\hbar\kappa \approx 0.1$~\si{\milli\electronvolt}), this Voigt profile differs from a Gaussian by less than 1\% in width, which justifies fitting the empty cavity envelope with a Gaussian to obtain the experimental reference $I_\text{mod,Ref}$. In the regime $\sigma \gg \kappa, \gamma+\gamma^*$ relevant to our experiment, the vibration distribution $P(\omega_a)$ is essentially constant over the narrow spectral range in which the emitter response varies, and the extinction of a single line probed by a co-polarised beam reduces to a Lorentzian (see Supplementary Note 3 for the derivation):
\begin{equation}
   1 - \frac{I_\text{mod}(\omega)}{I_\text{mod,Ref}(\omega)}= \frac{\mathcal{C}}{1+\mathcal{C}}\frac{1}{1+\left(\dfrac{\omega - \omega_X}{\dfrac{\gamma + \gamma^*}{2}\sqrt{1+\mathcal{C}}\ }\right)^2} .
   \label{eq:TmodsurTcav}
\end{equation}

Without polarisation analysis, each line produces a dip of the same FWHM and of about half this amplitude (Supplementary Note 1). Equation~\eqref{eq:TmodsurTcav} is independent of $\bar\omega_a$, so that the unpolarised extinction does not depend on the offset $\bar\omega_a-\omega_X$ in this regime. This is verified by numerical simulation of Eq.~\eqref{eq:Ttot_incoherent} over the full range of offsets explored experimentally (Supplementary Fig.~S6).

\subsection*{Extraction of the vacuum Rabi coupling $g$}
The total area of the two dips is computed from the two-Lorentzian fit [Eq.~\eqref{eq:dip_fitting}] as $\Lambda=\pi(a_1\Gamma_1+a_2\Gamma_2)/2$, and its relation to the Rabi coupling beyond the weak cooperativity limit is detailed in Supplementary Note 4. Representative parameters at mode $p=15$ are $\hbar\kappa=\SI{106}{\micro eV}$ and $\sigma\approx\SI{2.5}{meV}$ (FWHM $\approx\SI{6}{meV}$).

\section*{Supplementary information}
The article has accompanying supplementary information.

\section*{Acknowledgements}
This work was financially supported by the French National Research Agency (ANR) through the projects CSUPER2 (ANR-24-CE09-3415) and DELICACY (ANR-22-CE47-0001). The authors thank Pascal Morfin, Arnaud Leclercq, and the LPENS mechanical workshop for their assistance in designing and fabricating mechanical components. We also thank Aurélie Pierret and Mohamed-Raouf Amara for their help with the chemical synthesis at LPENS and Torben Pöpplau for assistance with fiber micromachining at LKB.

\section*{Author contributions}
Z.S. synthesized the pQDs, designed and implemented the microcavity setup under the supervision of Y.C, C.D., J.R. and C.V. The single pQD emission spectroscopy was performed by M.C.-T and Z.S., and the single pQD extinction spectroscopy by M.C.-T, all under the supervision of Y.C and C.D. The model was developed by Y.C. and the data analysis was done by M.C.-T, with the input of all coauthors. The paper was written by M.C.-T., Y.C. and C.D., with the contribution of all coauthors. Y.C. and C.D. conceived the project and supervised the work. All authors discussed the results and commented on the paper.

\section*{Competing interests}
The authors declare no competing interests.


\begin{thebibliography}{10}
\expandafter\ifx\csname url\endcsname\relax
  \def\url#1{\burl{#1}}\fi
\expandafter\ifx\csname urlprefix\endcsname\relax\def\urlprefix{URL }\fi
\providecommand{\bibinfo}[2]{#2}
\providecommand{\eprint}[2][]{\url{#2}}
\providecommand{\doi}[1]{\url{https://doi.org/#1}}
\bibcommenthead

\bibitem{Zumofen2008}
\bibinfo{author}{Zumofen, G.}, \bibinfo{author}{Mojarad, N.~M.},
  \bibinfo{author}{Sandoghdar, V.} \& \bibinfo{author}{Agio, M.}
\newblock \bibinfo{title}{Perfect reflection of light by an oscillating
  dipole}.
\newblock \emph{\bibinfo{journal}{Phys. Rev. Lett.}}
  \textbf{\bibinfo{volume}{101}}, \bibinfo{pages}{180404}
  (\bibinfo{year}{2008}).

\bibitem{Tey2008}
\bibinfo{author}{Tey, M.} \emph{et~al.}
\newblock \bibinfo{title}{Strong interaction between light and a single trapped
  atom without the need for a cavity.}
\newblock \emph{\bibinfo{journal}{Nature Physics}}
  \textbf{\bibinfo{volume}{4}}, \bibinfo{pages}{924–927}
  (\bibinfo{year}{2008}).

\bibitem{Vamivakas2007}
\bibinfo{author}{Vamivakas, A.~N.} \emph{et~al.}
\newblock \bibinfo{title}{Strong extinction of a far-field laser beam by a
  single quantum dot}.
\newblock \emph{\bibinfo{journal}{Nano Letters}} \textbf{\bibinfo{volume}{7}},
  \bibinfo{pages}{2892--2896} (\bibinfo{year}{2007}).

\bibitem{Pototschnig2011}
\bibinfo{author}{Pototschnig, M.} \emph{et~al.}
\newblock \bibinfo{title}{Controlling the phase of a light beam with a single
  molecule}.
\newblock \emph{\bibinfo{journal}{Phys. Rev. Lett.}}
  \textbf{\bibinfo{volume}{107}}, \bibinfo{pages}{063001}
  (\bibinfo{year}{2011}).

\bibitem{Tran2017}
\bibinfo{author}{Tran, T.~H.}, \bibinfo{author}{Siyushev, P.},
  \bibinfo{author}{Wrachtrup, J.} \& \bibinfo{author}{Gerhardt, I.}
\newblock \bibinfo{title}{Extinction of light and coherent scattering by a
  single nitrogen-vacancy center in diamond}.
\newblock \emph{\bibinfo{journal}{Phys. Rev. A}} \textbf{\bibinfo{volume}{95}},
  \bibinfo{pages}{053831} (\bibinfo{year}{2017}).

\bibitem{Wang2019}
\bibinfo{author}{Wang, D.} \emph{et~al.}
\newblock \bibinfo{title}{Turning a molecule into a coherent two-level quantum
  system}.
\newblock \emph{\bibinfo{journal}{Nature Physics}}
  \textbf{\bibinfo{volume}{15}}, \bibinfo{pages}{483–489}
  (\bibinfo{year}{2019}).

\bibitem{Tomm2024}
\bibinfo{author}{Tomm, N.} \emph{et~al.}
\newblock \bibinfo{title}{Realization of a coherent and efficient
  one-dimensional atom}.
\newblock \emph{\bibinfo{journal}{Phys. Rev. Lett.}}
  \textbf{\bibinfo{volume}{133}}, \bibinfo{pages}{083602}
  (\bibinfo{year}{2024}).

\bibitem{Codreanu2026}
\bibinfo{author}{Codreanu, N.} \emph{et~al.}
\newblock \bibinfo{title}{Above-unity coherent cooperativity of tin-vacancy
  centers in diamond photonic crystal cavities}.
\newblock \emph{\bibinfo{journal}{Phys. Rev. X}} \textbf{\bibinfo{volume}{16}},
  \bibinfo{pages}{021060} (\bibinfo{year}{2026}).

\bibitem{Koster2026}
\bibinfo{author}{Köster, K.} \emph{et~al.}
\newblock \bibinfo{title}{High-cooperativity coupling and spin-resolved
  extinction of tin-vacancy centers in a diamond-like microcavity}.
\newblock \emph{\bibinfo{journal}{PRX Quantum}}  (\bibinfo{year}{2026}).
\newblock \bibinfo{note}{In the press}.

\bibitem{Gruber2026}
\bibinfo{author}{Gruber, C.~G.} \emph{et~al.}
\newblock \bibinfo{title}{High-throughput in situ sizing and quantum yield
  determination of individual perovskite nanocrystals}.
\newblock \emph{\bibinfo{journal}{Nat. Mater.}} \textbf{\bibinfo{volume}{25}},
  \bibinfo{pages}{1344--1351} (\bibinfo{year}{2026}).

\bibitem{Raino2016}
\bibinfo{author}{Rain{\`o}, G.} \emph{et~al.}
\newblock \bibinfo{title}{Single cesium lead halide perovskite nanocrystals at
  low temperature: Fast single-photon emission, reduced blinking, and exciton
  fine structure}.
\newblock \emph{\bibinfo{journal}{ACS Nano}} \textbf{\bibinfo{volume}{10}},
  \bibinfo{pages}{2485--2490} (\bibinfo{year}{2016}).

\bibitem{Kaplan2023}
\bibinfo{author}{Kaplan, A. E.~K.} \emph{et~al.}
\newblock \bibinfo{title}{{{Hong-Ou-Mandel}} interference in colloidal
  {CsPbBr$_3$} perovskite nanocrystals}.
\newblock \emph{\bibinfo{journal}{Nat. Photonics}}
  \textbf{\bibinfo{volume}{17}}, \bibinfo{pages}{775--780}
  (\bibinfo{year}{2023}).

\bibitem{Jun2024}
\bibinfo{author}{Jun, S.} \emph{et~al.}
\newblock \bibinfo{title}{Ultrafast and bright quantum emitters from the
  cavity-coupled single perovskite nanocrystals}.
\newblock \emph{\bibinfo{journal}{ACS Nano}} \textbf{\bibinfo{volume}{18}},
  \bibinfo{pages}{1396--1403} (\bibinfo{year}{2024}).

\bibitem{Li2026}
\bibinfo{author}{Li, P.} \emph{et~al.}
\newblock \bibinfo{title}{Deterministic integration of {CsPbBr$_3$} quantum
  dots with plasmonic ring microcavities}.
\newblock \emph{\bibinfo{journal}{ACS Photonics}}
  \textbf{\bibinfo{volume}{13}}, \bibinfo{pages}{2006--2014}
  (\bibinfo{year}{2026}).

\bibitem{Liao2026}
\bibinfo{author}{Liao, T.-H.} \emph{et~al.}
\newblock \bibinfo{title}{Ultrafast, nonblinking single-photon sources from
  perovskite quantum dots in plasmonic nanocavities}.
\newblock \emph{\bibinfo{journal}{Science Advances}}
  \textbf{\bibinfo{volume}{12}}, \bibinfo{pages}{eaec4380}
  (\bibinfo{year}{2026}).

\bibitem{Oddi2026}
\bibinfo{author}{Oddi, V.} \emph{et~al.}
\newblock \bibinfo{title}{Purcell-enhanced single-photon generation from
  {CsPbBr$_3$} quantum dots in in situ selected {Laguerre}--{Gaussian} modes}.
\newblock \emph{\bibinfo{journal}{ACS Nano}} \textbf{\bibinfo{volume}{20}},
  \bibinfo{pages}{6167--6175} (\bibinfo{year}{2026}).

\bibitem{Said2026}
\bibinfo{author}{Said, Z.} \emph{et~al.}
\newblock \bibinfo{title}{Cavity {{Quantum Electrodynamics}} with {{Single
  Perovskite Quantum Dots}}: {{Assessing Rabi Coupling Strength}}, {{Pure
  Dephasing}}, and {{Spectral Diffusion}}}.
\newblock \emph{\bibinfo{journal}{Nano Letters}} \textbf{\bibinfo{volume}{26}},
  \bibinfo{pages}{4815--4821} (\bibinfo{year}{2026}).

\bibitem{Amara2023}
\bibinfo{author}{Amara, M.-R.} \emph{et~al.}
\newblock \bibinfo{title}{Spectral {Fingerprint} of {Quantum} {Confinement} in
  {Single} {CsPbBr}$_{\textrm{3}}$ {Nanocrystals}}.
\newblock \emph{\bibinfo{journal}{Nano Letters}} \textbf{\bibinfo{volume}{23}},
  \bibinfo{pages}{3607--3613} (\bibinfo{year}{2023}).

\bibitem{Gardiner1985}
\bibinfo{author}{Gardiner, C.~W.} \& \bibinfo{author}{Collett, M.~J.}
\newblock \bibinfo{title}{Input and output in damped quantum systems: Quantum
  stochastic differential equations and the master equation}.
\newblock \emph{\bibinfo{journal}{Phys. Rev. A}} \textbf{\bibinfo{volume}{31}},
  \bibinfo{pages}{3761--3774} (\bibinfo{year}{1985}).

\bibitem{Auffeves2010}
\bibinfo{author}{Auff\`eves, A.} \emph{et~al.}
\newblock \bibinfo{title}{Controlling the dynamics of a coupled atom-cavity
  system by pure dephasing}.
\newblock \emph{\bibinfo{journal}{Phys. Rev. B}} \textbf{\bibinfo{volume}{81}},
  \bibinfo{pages}{245419} (\bibinfo{year}{2010}).

\bibitem{Auffeves2007}
\bibinfo{author}{Auff\`eves-Garnier, A.}, \bibinfo{author}{Simon, C.},
  \bibinfo{author}{G\'erard, J.-M.} \& \bibinfo{author}{Poizat, J.-P.}
\newblock \bibinfo{title}{Giant optical nonlinearity induced by a single
  two-level system interacting with a cavity in the {{Purcell}} regime}.
\newblock \emph{\bibinfo{journal}{Phys. Rev. A}} \textbf{\bibinfo{volume}{75}},
  \bibinfo{pages}{053823} (\bibinfo{year}{2007}).

\bibitem{Borel2023}
\bibinfo{author}{Borel, A.} \emph{et~al.}
\newblock \bibinfo{title}{Telecom band single-photon source using a grafted
  carbon nanotube coupled to a fiber {Fabry–Perot} cavity in the {Purcell}
  regime}.
\newblock \emph{\bibinfo{journal}{ACS Photonics}}
  \textbf{\bibinfo{volume}{10}}, \bibinfo{pages}{2839--2845}
  (\bibinfo{year}{2023}).

\bibitem{Najer2019}
\bibinfo{author}{Najer, D.} \emph{et~al.}
\newblock \bibinfo{title}{A gated quantum dot strongly coupled to an optical
  microcavity}.
\newblock \emph{\bibinfo{journal}{Nature}} \textbf{\bibinfo{volume}{575}},
  \bibinfo{pages}{622--627} (\bibinfo{year}{2019}).

\bibitem{Zhu2024}
\bibinfo{author}{Zhu, C.} \emph{et~al.}
\newblock \bibinfo{title}{Single-photon superradiance in individual caesium
  lead halide quantum dots}.
\newblock \emph{\bibinfo{journal}{Nature}} \textbf{\bibinfo{volume}{626}},
  \bibinfo{pages}{535--541} (\bibinfo{year}{2024}).

\bibitem{Boehme2025}
\bibinfo{author}{Boehme, S.~C.} \emph{et~al.}
\newblock \bibinfo{title}{Single-photon superabsorption in {CsPbBr$_3$}
  perovskite quantum dots}.
\newblock \emph{\bibinfo{journal}{Nat. Photonics}}
  \textbf{\bibinfo{volume}{19}}, \bibinfo{pages}{864--870}
  (\bibinfo{year}{2025}).

\bibitem{Sun2023}
\bibinfo{author}{Sun, W.} \emph{et~al.}
\newblock \bibinfo{title}{Elastic phonon scattering dominates dephasing in
  weakly confined cesium lead bromide nanocrystals at cryogenic temperatures}.
\newblock \emph{\bibinfo{journal}{Nano Lett.}} \textbf{\bibinfo{volume}{23}},
  \bibinfo{pages}{2615--2622} (\bibinfo{year}{2023}).

\bibitem{Protesescu2015}
\bibinfo{author}{Protesescu, L.} \emph{et~al.}
\newblock \bibinfo{title}{Nanocrystals of cesium lead halide perovskites
  ({CsPbX}$_{\textrm{3}}$, {X = Cl}, {Br}, and {I}): Novel optoelectronic
  materials showing bright emission with wide color gamut}.
\newblock \emph{\bibinfo{journal}{Nano Letters}} \textbf{\bibinfo{volume}{15}},
  \bibinfo{pages}{3692--3696} (\bibinfo{year}{2015}).

\end{thebibliography}
\end{document}

\title[Article Title]{Supplementary Information: Extinction spectroscopy of a single colloidal perovskite quantum dot coupled to a tunable microcavity}

\author[1]{\fnm{Marina} \sur{Cagnon-Trouche}}

\author[1]{\fnm{Zakaria} \sur{Said}}

\author[2]{\fnm{Jakob} \sur{Reichel}}

\author[1]{\fnm{Christophe} \sur{Voisin}}

\author*[1]{\fnm{Yannick} \sur{Chassagneux}}
\email{yannick.chassagneux@phys.ens.fr}

\author*[1,3]{\fnm{Carole} \sur{Diederichs}}
\email{carole.diederichs@phys.ens.fr}

\affil[1]{\orgdiv{Laboratoire de Physique de l’Ecole normale supérieure}, \orgname{ENS, Université PSL, CNRS, Sorbonne Université, Université Paris Cité}, \orgaddress{\street{24 rue Lhomond}, \city{Paris}, \postcode{F-75005}, \country{France}}}

\affil[2]{\orgdiv{Laboratoire Kastler Brossel}, \orgname{Sorbonne Université, CNRS, ENS - Université PSL, Collège de France}, \orgaddress{\city{Paris} \postcode{F-75252},  \country{France}}}

\affil[3]{\orgname{Institut Universitaire de France}, \orgaddress{\city{Paris}, \postcode{75231}, \country{France}}}

\maketitle
%\doublespacing
%\linenumbers

\section*{Supplementary Note 1: Polarisation-insensitive detection}

\begin{figure}
    \centering
    \includegraphics[width=0.8\linewidth]{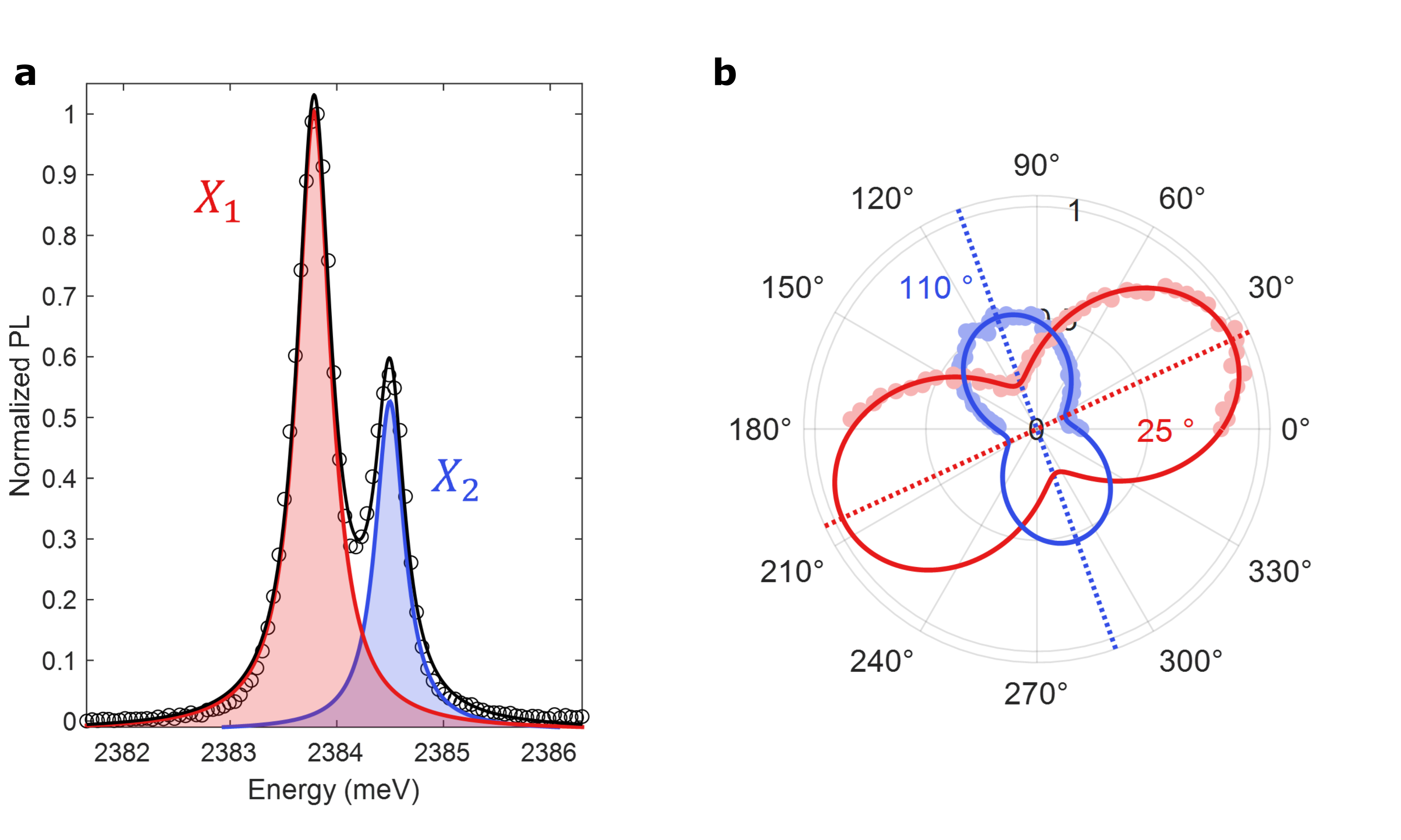}
    \caption{\textbf{Characterisation of PL in free space.} (a) Free-space PL spectrum at 10 K, showing the fine structure of the exciton with two linearly and orthogonally polarised lines, $X_1$ in red and $X_2$ in blue. (b) Polarisation diagram: integrated intensity of the $X_1$ and $X_2$ lines as a function of the analyser angle, fitted to Malus’s law (solid lines).}
    \label{fig:spectrum_PL_pol}
\end{figure}

The exciton fine structure consists of two lines, $X_1$ and $X_2$, linearly polarised along orthogonal directions (Fig.~\ref{fig:spectrum_PL_pol}), which we model as a pair of orthogonal dipoles. The cavity supports two orthogonally polarised modes, whose splitting is small compared to the cavity linewidth, so that they can be considered degenerate. Choosing the polarisation basis of the modes along the dipole axes, each line couples only to its co-polarised mode. As the two modes are orthogonally polarised, they do not interfere, and without polarisation analysis the measured modulated intensities are the sums of the intensities transmitted in the two polarisations, both with and without the emitter (Fig.~1 of the main text): 
\begin{equation}
I_\mathrm{tot}=I_1+I_2,
\qquad
I_\mathrm{Ref}=I_\mathrm{Ref1}+I_\mathrm{Ref2}.
\end{equation}
The measured modulated transmission is given by:
\begin{equation}
T=\frac{I_\mathrm{tot}}{I_\mathrm{Ref}}=\frac{I_\mathrm{Ref1}}{I_\mathrm{Ref}}\frac{I_1}{I_\mathrm{Ref1}}+\frac{I_\mathrm{Ref2}}{I_\mathrm{Ref}}\frac{I_2}{I_\mathrm{Ref2}}=\frac{I_\mathrm{Ref1}}{I_\mathrm{Ref}}T_1+\frac{I_\mathrm{Ref2}}{I_\mathrm{Ref}}T_2.
\end{equation}

For equal reference intensities in the two polarisations ($I_\mathrm{Ref1}=I_\mathrm{Ref2}=\frac{1}{2}I_\mathrm{Ref}$), the modulated transmission is then the average of the two polarisation-resolved transmissions $T_1$ and $T_2$:
\begin{equation}
T=\frac{1}{2}\left(T_1+T_2\right).
\label{eq:T_mean}
\end{equation}
If the reference is not perfectly balanced between the two polarisations, each polarisation-resolved transmission is weighted by the fraction $I_{\mathrm{Ref}i}/I_\mathrm{Ref}$ instead of one half. We consider in the following the balanced case expected for an unpolarised probe.

Fig.~\ref{fig:simulation_deux_raies} illustrates this situation, computed with the model of Supplementary Note~3 for parameters of the same order as those of mode $p=15$: an extinction of 11.2\% per line for a co-polarised probe appears as dips of 5.9\% in the measured transmission. This is slightly more than half because the transmission minimum at each line also contains the tail of the other line. The amplitudes of the two-Lorentzian fit [Eq.~(5) of the main text] remove this contribution, and are used to quantify the unpolarised extinction of each line.

% Compute_2emitter_mieux_v2_SI.m
 \begin{figure}[H]
   \centering
   \includegraphics{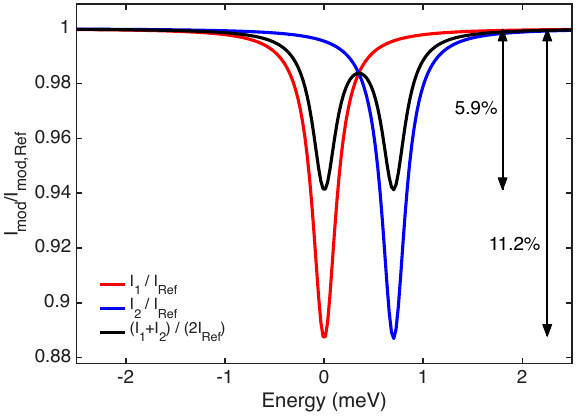}
    \caption{\textbf{Effect of polarisation-insensitive detection on the extinction dips.} Numerical simulation of the normalised transmission for the two fine structure lines of the exciton. The red ($T_1$) and blue ($T_2$) curves correspond to a probe polarised along the line of interest, which couples only to the co-polarised cavity mode. The black curve ($T$) corresponds to our experimental configuration, with polarisation-insensitive excitation and detection: the reference then contains both cavity modes, and each dip is about half as deep as for a co-polarised probe (5.9\% instead of 11.2\%), slightly more than half because of the overlap with the tail of the other line. Parameters: $\hbar\kappa=\SI{106}{\micro eV}$, $\hbar g=\SI{30}{\micro eV}$ for each line, $\hbar(\gamma+\gamma^*)=\SI{270}{\micro eV}$ and $\sigma=\SI{2.5}{meV}$ (FWHM \SI{6}{meV}).}
   \label{fig:simulation_deux_raies}
 \end{figure}

\section*{Supplementary Note 2: Additional transmission data}

\begin{figure}[H]
   \centering
   \includegraphics{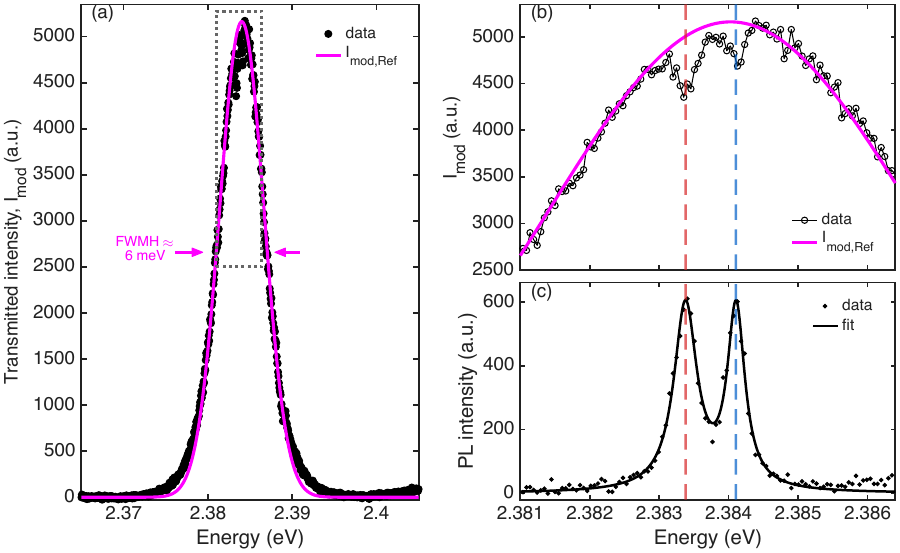}
   \caption{\textbf{Transmission spectrum at mode $p=17$.} (a) Transmitted intensity $I_\mathrm{mod}$ of the cavity tuned to the pQD (black dots), and baseline $I_\mathrm{mod,Ref}$ obtained from a Gaussian fit excluding a 2-meV wide window centred on excitonic doublet (magenta). The exciton lines lie close to the centre of the vibration distribution. The dotted rectangle delimits the region magnified in (b). (b) Magnified view of (a) around the exciton energy, showing the two extinction dips. (c) Photoluminescence spectrum of the pQD coupled to the cavity (dots) and fit (line). In (b) and (c), the dashed lines indicate the energies of $X_1$ (red) and $X_2$ (blue).}
   \label{fig:extraction_T0}
 \end{figure}

Figure~\ref{fig:extraction_T0} shows a transmission spectrum recorded at mode $p=17$ with the exciton lines close to the centre of the vibration distribution, complementary to the spectrum of Fig.~2 of the main text, where they lie on its flank. In both cases, the two dips appear at the energies of the $X_1$ and $X_2$ lines measured by photoluminescence in cavity.

% Plot_baseline_gauss_exemple_SI.m
 
\begin{figure}[H]
   \centering
   \includegraphics{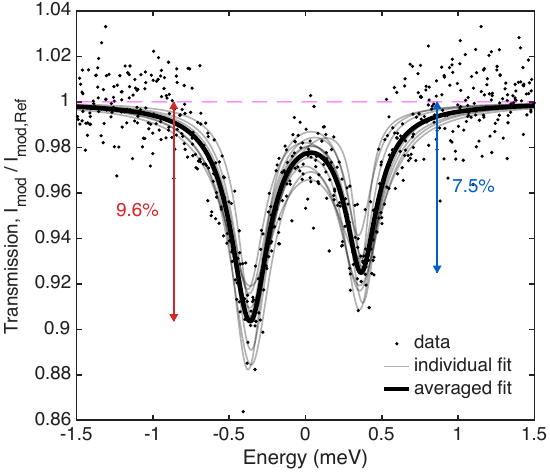}
   \caption{\textbf{Reproducibility of the extinction dips at a given mode.} Normalised transmission $I_\mathrm{mod}/I_\mathrm{mod,Ref}$ at mode $p=15$ for ten acquisitions recorded on different days (dots), with the individual Lorentzian fits (grey lines) and their average (black line). The arrows indicate the mean depths at the transmission minima, $9.6\pm0.4\%$ for $X_1$ and $7.5\pm0.3\%$ for $X_2$.}
   \label{fig:creux_mode15}
 \end{figure}

 Figure~\ref{fig:creux_mode15} gathers the ten transmission spectra recorded at mode $p=15$ on different days, illustrating the reproducibility of the unpolarised extinctions. Each spectrum is fitted by a sum of two Lorentzians [Eq.~(5) of the main text]. The mean depths at the transmission minima are $9.6\pm0.4\%$ for $X_1$ and $7.5\pm0.3\%$ for $X_2$, and the mean unpolarised extinctions, given by the Lorentzian amplitudes, are 9.5\% and 7.1\%. 

% plot_allmean_fit_pour1mode_T_T0_SI.m

\begin{figure}[H]
   \centering
   \includegraphics{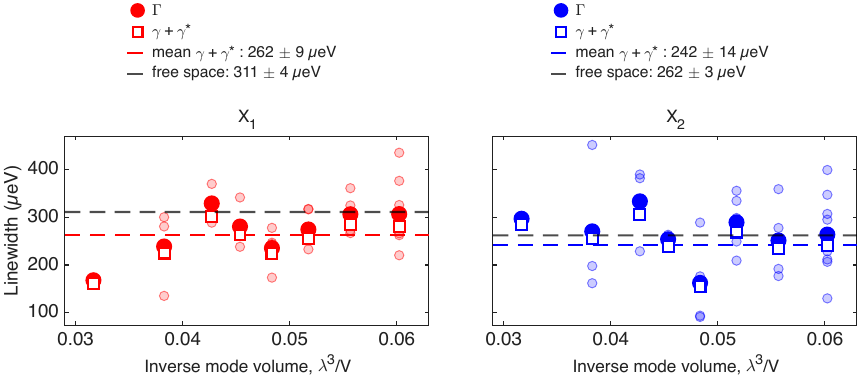}
   \caption{\textbf{Extinction dip linewidths versus inverse mode volume.} Full width at half maximum (FWHM) $\Gamma$ of the extinction dips of $X_1$ (red) and $X_2$ (blue), extracted from the Lorentzian fits of the 33 acquisitions recorded across all longitudinal modes, as a function of the inverse mode volume $\lambda^3/V$. Transparent dots are individual acquisitions and full circles their mean for each mode. Squares are the means corrected for the cavity-induced broadening, $\Gamma/\sqrt{1+\mathcal{C}}$, which gives $\gamma + \gamma ^*$. The coloured dashed lines indicate the mean corrected linewidths, and the black dashed lines the mean free-space photoluminescence linewidths.}
   \label{fig:linewitdh_ext}
\end{figure}

Figure~\ref{fig:linewitdh_ext} shows the dip linewidths $\Gamma$ extracted from all 33 acquisitions as a function of the inverse mode volume. They show a large dispersion and no systematic dependence on the mode volume. Coupling to the cavity is expected to broaden each dip as $\Gamma = (\gamma+\gamma^*) \sqrt{1+\mathcal{C}}$ (Supplementary Note~3), i.e. by up to about 10\% at the smallest mode volume. For each mode, we therefore divide the mean FWHM by $\sqrt{1+\mathcal{C}}$, where $\mathcal{C}$ is the mean cooperativity deduced from the dip area at this mode (Supplementary Note~4). Averaging over all modes yields corrected linewidths of $262\pm9$~\si{\micro eV} for $X_1$ and $242\pm14$~\si{\micro eV} for $X_2$, both consistent within their uncertainties with the instantaneous linewidth of 250~\si{\micro eV} measured independently on this pQD \cite{Said2026}. Compared with the free space photoluminescence linewidths of the same pQD, $311\pm4$~\si{\micro eV} and $262\pm3$~\si{\micro eV}, the extinction lines are narrower by 16\% for $X_1$, a statistically significant difference ($t_s\simeq 4.98$), and by 8\% for $X_2$, a difference which is not significant ($t_s\simeq 1.40$).

\section*{Supplementary Note 3: Model of the modulated transmission}

\begin{figure}[H]
   \centering
   \includegraphics{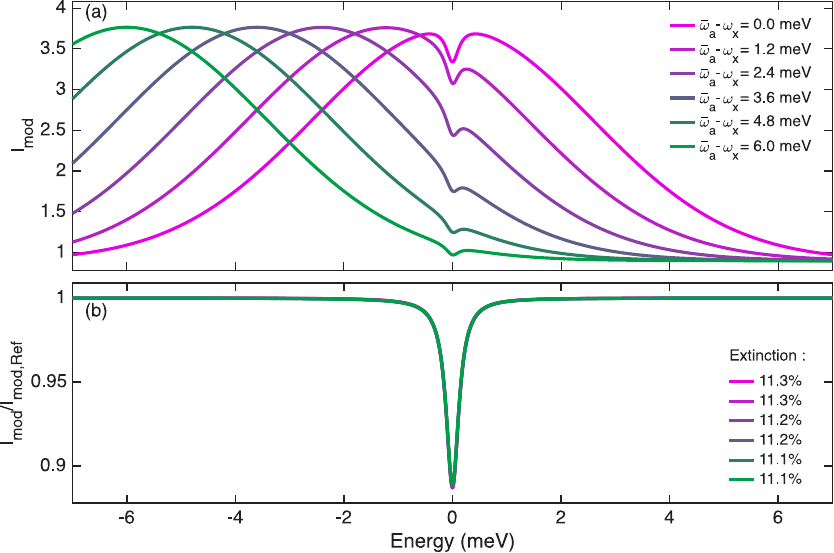}
    \caption{\textbf{Independence of the dip depth from the position of the emitter within the vibration distribution.} Numerical simulation of (a) the modulated transmitted intensity $I_\mathrm{mod}$ and (b) the normalised transmission $I_\mathrm{mod}/I_\mathrm{mod,Ref}$ for a single line probed with a co-polarised beam, for offsets $\bar\omega_a-\omega_X$ between the mean cavity frequency and the emitter frequency ranging from 0 to \SI{6}{meV}, i.e. up to the FWHM of the vibration distribution. The dip depth, given in the legend of (b), varies by less than 0.2 percentage points. Parameters: $\hbar\kappa=\SI{106}{\micro eV}$, $\hbar g=\SI{30}{\micro eV}$, $\hbar(\gamma+\gamma^*)=\SI{270}{\micro eV}$ and $\sigma=\SI{2.5}{meV}$ (FWHM \SI{6}{meV}).}
   \label{fig:Tmod_delta}
 \end{figure}

We denote by $L_{\Gamma,\omega_0}$ the normalised Lorentzian of FWHM $\Gamma$ centred at $\omega_0$, by $G_{\sigma,\omega_0}$ the normalised Gaussian of standard deviation $\sigma$ centred at $\omega_0$, and by $V_{\Gamma,\sigma,\omega_0}=L_{\Gamma,\omega_0}*G_{\sigma,0}$ their convolution, the Voigt profile. The distribution of the cavity frequency is $P(\omega_a)=G_{\sigma,\bar\omega_a}(\omega_a)$ [Eq.~(6) of the main text], and the intensity transmitted in the polarisation of each line is $I_\mathrm{mod}(\omega)=\int|t(\omega,\omega_a)|^2P(\omega_a)\,\mathrm{d}\omega_a$ [Eq.~(2) of the main text]. To compute this integral, we separate the real and imaginary parts of the denominator of the transmission amplitude [Eq.~(1) of the main text]:
\begin{equation}
t(\omega,\omega_a)=\frac{\kappa/2}{A_r(\omega)+i\,A_i(\omega,\omega_a)},
\end{equation}
with
\begin{equation}
A_r(\omega)=\frac{\kappa}{2}+\frac{g^2\,\dfrac{\gamma+\gamma^*}{2}}{\left(\dfrac{\gamma+\gamma^*}{2}\right)^2+(\omega-\omega_X)^2},
\qquad
A_i(\omega,\omega_a)=-(\omega-\omega_a)+\frac{g^2\,(\omega-\omega_X)}{\left(\dfrac{\gamma+\gamma^*}{2}\right)^2+(\omega-\omega_X)^2}.
\end{equation}
$A_r$ does not depend on $\omega_a$. At fixed $\omega$, $|t|^2$ is thus proportional to a Lorentzian of $A_i$, and therefore of $\omega_a$:
\begin{equation}
|t(\omega,\omega_a)|^2=\pi\,\frac{(\kappa/2)^2}{A_r(\omega)}\,L_{2A_r(\omega),0}\big(A_i(\omega,\omega_a)\big),
\end{equation}
and its average over $P(\omega_a)$ is a Voigt profile, i.e.\ the convolution of a Lorentzian of width $2A_r$, set by the losses of the system, with a Gaussian of standard deviation $\sigma$, set by the vibration amplitude:
\begin{equation}
I_\mathrm{mod}(\omega)=\pi\,\frac{(\kappa/2)^2}{A_r(\omega)}\,V_{2A_r(\omega),\sigma,0}\big(A_i(\omega,\bar\omega_a)\big).
\end{equation}
For an empty cavity ($g=0$), this expression reduces to $I_\mathrm{mod,Ref}(\omega)=\pi\,\frac{\kappa}{2}\,V_{\kappa,\sigma,\bar\omega_a}(\omega)$. Since $\kappa\ll\sigma$, this Voigt profile is close to a Gaussian, which justifies the Gaussian fit used to obtain the experimental reference. When the vibration amplitude dominates all spectral widths ($\sigma\gg\kappa,\gamma+\gamma^*$), the Gaussian varies little over the Lorentzian, and
\begin{equation}
I_\mathrm{mod}(\omega)\simeq\pi\left(\frac{\kappa}{2}\right)^{2}\frac{P(\omega)}{A_r(\omega)},
\qquad
I_\mathrm{mod,Ref}(\omega)\simeq\pi\,\frac{\kappa}{2}\,P(\omega).
\end{equation}
The vibration distribution then cancels in the normalised transmission, $1-I_\mathrm{mod}/I_\mathrm{mod,Ref}=1-(\kappa/2)/A_r(\omega)$, which yields the Lorentzian of Eq.~(7) of the main text, of amplitude $\mathcal{C}/(1+\mathcal{C})$ and FWHM $(\gamma+\gamma^*)\sqrt{1+\mathcal{C}}$ for a co-polarised probe. This expression does not depend on $\bar\omega_a$. Fig.~\ref{fig:Tmod_delta} confirms with the full model [Eq.~(2) of the main text] that the dip depth is unchanged when the offset $\bar\omega_a-\omega_X$ between the mean cavity frequency and the emitter frequency varies. This point is important for the analysis, since this offset can vary from one acquisition to the next.

%% compute_Tmod_Pa_SI.m

\section*{Supplementary Note 4: Extraction of the vacuum Rabi coupling}

The extinction of line $X_i$ for a co-polarised probe, $1-T_i$, is given by the Lorentzian [Eq.~(7) of the main text]:

\begin{equation}
\frac{\mathcal{C}_i}{1+\mathcal{C}_i}\frac{1}{1+\left(\dfrac{\omega - \omega_{X_i}}{\dfrac{\gamma_i + \gamma_i^*}{2}\sqrt{1+\mathcal{C}_i}\ }\right)^2} ,
   \label{eq:TmodsurTcav_i}
\end{equation}
where $\omega_{X_i}$, $\hbar\left(\gamma_i+\gamma_i^*\right)$ and $\mathcal{C}_i=4 g_i^2/[\kappa (\gamma_i+\gamma_i^*)]$ are the frequency, instantaneous linewidth and cooperativity of line $X_i$, and $g_i$ its vacuum Rabi coupling. From Eq.~\eqref{eq:T_mean}, the measured extinction is 
\begin{equation}
1-T=\frac{1}{2}\Big[\left(1-T_1\right)+\left(1-T_2\right)\Big].
\label{eq:ext_mean}
\end{equation}
Integrating Eq.~\eqref{eq:ext_mean} over the whole spectrum gives the total area of the two dips,
\begin{equation}
    \Lambda=\int\left(1-T\right)\mathrm{d}\omega =\frac{1}{2}\sum_{i=1,2}\pi\,\frac{\gamma_i+\gamma_i^*}{2}\,\frac{\mathcal{C}_i}{\sqrt{1+\mathcal{C}_i}}.
\end{equation}
In the weak cooperativity limit ($\mathcal{C}_i\ll1$), each term reduces to $\pi g_i^2/\kappa$, independently of the instantaneous linewidth, so that
\begin{equation}
\Lambda\simeq\frac{2\pi}{\kappa}\,\langle g^2\rangle,
\qquad
\langle g^2\rangle=\frac{g_1^2+g_2^2}{2}.
\end{equation}
Experimentally, the area is obtained from the two-Lorentzian fit [Eq.~(5) of the main text] as $\Lambda = \frac{\pi}{2} \sum_{i=1,2} a_i \Gamma_i$, and Eq.~(4) of the main text thus yields the mean squared vacuum Rabi coupling of the two lines, $\hbar^2 \langle g^2 \rangle$. Assuming the same instantaneous linewidth for both lines, the mean cooperativity is then $\langle \mathcal{C} \rangle = 4\langle g^2 \rangle /[\kappa (\gamma+\gamma^*)]$, which reaches about 0.2 at the smallest mode volume. This mean cooperativity is used to correct the linewidths for the cavity-induced broadening (Supplementary Note~2).

\bibliography{Bibliography}